\documentstyle[twoside,fleqn,espcrc2,amssymb]{article}

\newcommand{\dd}{{\rm d}}

\newcommand{\eq}{\begin{equation}}
\newcommand{\feq}{\end{equation}}
\newcommand{\arr}{\begin{eqnarray}}
\newcommand{\farr}{\end{eqnarray}}

\newcommand{\real}{\relax{\rm I\kern-.18em R}}

\newcommand{\AmS}{{\protect\the\textfont2
  A\kern-.1667em\lower.5ex\hbox{M}\kern-.125emS}}

\title{Monopole Condensation and Color Confinement}

\author{A. Di Giacomo\address{Dipartimento
        di Fisica and INFN, Universit\`a di Pisa, I-56100 Pisa, Italy},
	B. Lucini\address{Scuola Normale Superiore and INFN 
			sezione di Pisa, I-56100 Pisa, Italy},
        M. Montesi\thanks{Speaker at the conference.}$^{\rm b}$
and 	G. Paffuti$^{\rm a}$
}

\begin{document}

\begin{abstract}
New evidence is discussed of monopole condensation in the vacuum of $SU(2)$ 
and $SU(3)$ gauge theories.
Monopoles defined by different abelian projections do condense in the 
transition to the confined phase and show the same behavior.
For $SU(2)$ critical indices are determined by finite size scaling 
analysis and the results agree with the 3d Ising Model, as expected.
\end{abstract}

\maketitle

\section{INTRODUCTION}

A particularly interesting attempt to understand the structure of the QCD
vacuum and color confinement is the dual superconductor mechanism
\cite{thoo75,mand76}.
By dual Meissner effect, in the presence of a quark-antiquark pair
the chromoelectric field is squeezed into flux tubes and the potential energy
rises linearly with the distance producing confinement.

The detection on the lattice of linear rising potential \cite{creu80} 
and of flux tube 
configurations \cite{haym87} 
has given support to this idea.
Direct evidence may come from the study of the symmetry of QCD vacuum.
Dual superconductivity is indeed the spontaneous 
breaking of the $U(1)$ symmetry related to monopole charge conservation. 
The color deconfining transition occurs between the broken
magnetic phase, where magnetic monopoles condense, and 
the symmetric phase where the vacuum is invariant under the
action of U(1) magnetic group.

This order-disorder transition can be studied by constructing an operator 
with non trivial magnetic charge and computing its vacuum expectation value
({\it v.e.v.}) \cite{digi97}.
A creation operator for a monopole can be defined
and its {\it v.e.v.} used as an order parameter.

In non abelian gauge theories monopoles can be defined by means 
of abelian projection \cite{thoo81} in analogy with
the `t Hooft-Polyakov monopole \cite{thoo74,poly74} in the
Georgi-Glashow model.
In case of $SU(2)$ or $SU(3)$ pure gauge theories, 
as well as QCD, there is no
fundamental Higgs fields, but any operator $\Phi(x)$ in the adjoint 
representation of the gauge group can play its role.

The abelian projection is set diagonalizing the $\Phi$ operator. Under the residual $U(1)^{N-1}$ gauge subgroup the diagonal components of the vector field
transform as abelian fields.
There are abelian monopoles in the sites
where $\Phi(x)$ has two equal eigenvalues and 
the diagonalization is singular.
In $SU(2)$ gauge theories this means that
$\vec{\Phi}(x)\cdot \sigma$ is zero.
It is worth underlining that monopole charge is color singlet and gauge
invariant. Condensation is no breaking of the gauge symmetry group.

A priori on the lattice parallel transport on any 
closed path  can be chosen as
$\Phi(x)$. There are an infinite number of possible choices, for example 
the open plaquette $U_{XY}(x)$, the open Polyakov loop $P(x)$ and the 
``butterfly'' $U_{XY}U_{ZT}$.
The question is:
What kind of monopoles do condense if any?
Even if on a single configuration the number and the location of monopoles 
depend on the choice of the abelian projection, `t Hooft \cite{thoo81} 
guessed that 
different monopoles are physically
equivalent on the average and do condense in the QCD vacuum.

The creation operator of a monopole
$\mu(\vec{y},t)$ shifts field configurations 
$|a_{\mu}(\vec{x},t)\rangle$
by a field $a_{\mu}^{(top)}(\vec{x}-\vec{y},t)$ with non trivial 
magnetic charge (for example the Dirac potential)
\newpage
\eq
\mu(\vec{y},t)|a_{\mu}(\vec{x},t)\rangle = |a_{\mu}(\vec{x},t) 
+ a_{\mu}^{(top)}(\vec{x}-\vec{y},t)\rangle \ .
\feq

The original construction goes back to \cite{kada71} and has been developed
in different forms and different models by many authors
\cite{digi97,mari82,froh87,dario98}.
Dealing with compact gauge groups technical modification are needed due 
to the compact nature of field variables that cannot be shifted arbitrary 
\cite{digi97}. For details about non abelian gauge theories
 \cite{digi98,biagio98}.
The {\it v.e.v} of $\mu$ is given by
\eq
\langle\mu(\vec{y},t)\rangle = \frac{Z[S_m]}{Z[S]}
\feq
where $S$ is the usual action of the system and $S_m$ is the action modified by the addition of a monopole.

\section{RESULTS}

Instead of $\langle \mu \rangle$ it is numerically convenient to measure:
\eq
\rho = \frac{\dd}{\dd \beta} \log \langle\mu\rangle = \langle S\rangle_S 
- \langle S_m \rangle _{S_m} 
\feq
and since $\langle \mu\rangle = 1$ for $\beta = 0$ 
\eq
\langle\mu\rangle = \exp\int_0^{\beta} \rho(\beta')\dd\beta' \ .
\feq

\begin{figure}[htb]
\vspace{5cm}
\includegraphics{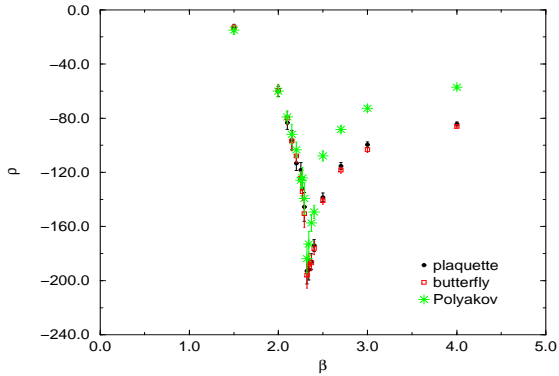}
\vspace{-1.3cm}
\caption{$\rho$ vs. $\beta$ for different abelian projections in $SU(2)$.}
\vspace{-0.5cm}
\label{fig1}
\end{figure}

In fig.~\ref{fig1} we show the typical behavior of $\rho$ vs. $\beta$ 
for 
$SU(2)$ gauge theory and for different abelian projections.
 These data have been obtained on a Quadrics Machine.

Our main results are:

1) Different abelian projections show the same behavior.

2) Enlarging the spatial size of the lattice (at fixed temporal size)
$\rho$ remains finite for $\beta < \beta_c$, i.e. 
 $\mu$ remains different from zero in the infinite volume limit. 
The asymptotic value at large $\beta$'s diverges to $-\infty$ 
with the spatial dimension, so that
$\mu$ goes to zero in the infinite volume limit in the 
deconfined phase $\beta > \beta_c$.

\begin{figure}[htb]
\vspace{5cm}
\includegraphics{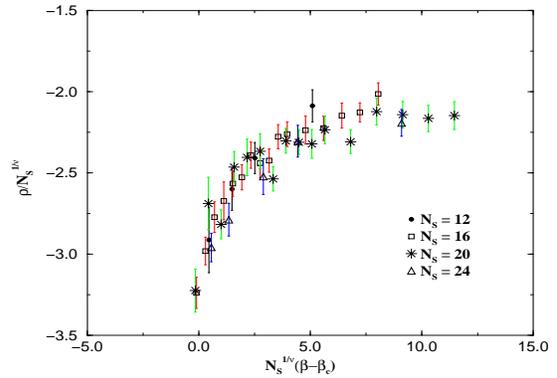}
\vspace{-1.3cm}
\caption{Finite size rescaling of data according eq.~\ref{fsc} with $\nu = 0.63$. Plaquette projection.}
\vspace{-0.5cm}
\label{fig2}
\end{figure}
Finite size scaling requires that in the critical region 
$\beta \simeq \beta_c$
\eq\label{fsc}
\rho/ L^{1/\nu} = f(L^{1/\nu}(\beta - \beta_c))
\feq
with $\nu$ the critical index related to the correlation length
\eq
\xi \sim (\beta_c - \beta)^{-\nu} \ .
\feq

How well scaling is obeyed is shown in fig.~\ref{fig2} where
data are rescaled according to eq.~\ref{fsc}
with $\nu = 0.63$ and $\beta_c = 2.295$ \cite{kars93}.
This is in agreement with the expectation that $SU(2)$ presents a phase 
transition of second order and belongs to the same universality class as 3d 
Ising model, where $\nu = 0.631(1)$.
\begin{figure}[htb]
\vspace{5cm}
\includegraphics{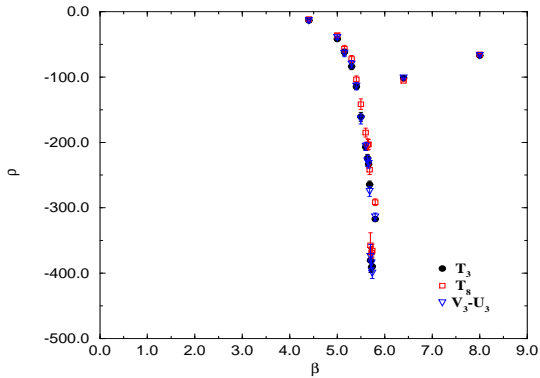}
\vspace{-1.3cm}
\caption{$\rho$ for different monopoles of the same abelian projection
 in $SU(3)$.
 Polyakov projection.}
\vspace{-0.5cm}
\label{fig3}
\end{figure}
\begin{figure}[htb]
\vspace{5cm}
\includegraphics{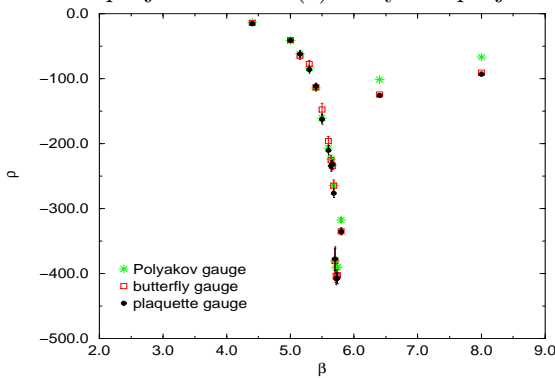}
\vspace{-1.3cm}
\caption{$\rho$ for different abelian projections in $SU(3)$.}
\vspace{-0.5cm} 
\label{fig4}
\end{figure}
%

3) The same analysis can be done in $SU(3)$.
In this case there are two independent monopoles related to the two 
independent diagonal generators of SU(3) group
for a given abelian projection. 
The corresponding $\rho$'s are plotted
for the Polyakov projection in fig.~\ref{fig3}
and coincide within errors.
Like for $SU(2)$, 
there is no difference between different abelian projections (Polyakov, spatial plaquette and butterfly) (see fig~\ref{fig4}).

The finite size scaling analysis for $SU(3)$ is on the way.
The expectation is $\nu = 1/3$ the phase transition being of first order.

\section{CONCLUSIONS}

Studying the symmetry properties of $SU(2)$ and $SU(3)$ gauge theories by 
means of a disorder parameter 
we find evidence that monopoles do condense in the confined phase.

Monopole condensation takes place in all the abelian projections 
we have studied. 
This shows that QCD vacuum
is more complicated than a simple $U(1)$ superconductor.

In $SU(2)$ the scaling behavior of the disorder parameter 
gives a critical index equal to that of 3d Ising model.

\section{Acknowledgements}

Partially supported by MURST, project: ``Fisica Teorica delle Interazioni 
Fondamentali''.
A. Di Giacomo acknowledges the financial contribution of the 
European Commission under the TMR-Program ERBFMRX-CT97-0122.

\end{document}